%% file: main.tex
\documentclass[conference,a4paper]{IEEEtran}
\usepackage{mdframed} 
\usepackage{cite}
\usepackage{amsmath}
\usepackage{graphicx}
\usepackage{nomencl}
\makenomenclature
\usepackage{etoolbox}
\renewcommand\nomgroup[1]{%
  \item[\bfseries
  \ifstrequal{#1}{P}{Parameters}{%
  \ifstrequal{#1}{V}{Variables}{%
  \ifstrequal{#1}{W}{Other Symbols}{}}}%
]}
\ifCLASSINFOpdf
\else
\fi
\begin{document}
%
\title{Optimal Real-time Coordination of Distributed Energy Resources in Low-voltage Grids}



%
\author{\IEEEauthorblockN{Sen Zhan\IEEEauthorrefmark{1},
Johan Morren\IEEEauthorrefmark{1}\IEEEauthorrefmark{2},
Wouter van den Akker\IEEEauthorrefmark{1}\IEEEauthorrefmark{3}, 
Anne van der Molen\IEEEauthorrefmark{1}\IEEEauthorrefmark{4}, and
Han Slootweg\IEEEauthorrefmark{1}\IEEEauthorrefmark{2}}
\IEEEauthorblockA{\IEEEauthorrefmark{1}Department of Electrical Engineering\\ Eindhoven University of Technology, Eindhoven, the Netherlands}
\IEEEauthorblockA{\IEEEauthorrefmark{2}Enexis Netbeheer, 's-Hertogenbosch, the Netherlands}
\IEEEauthorblockA{\IEEEauthorrefmark{3}Liander NV, Arnhem, the Netherlands}
\IEEEauthorblockA{\IEEEauthorrefmark{4}Stedin Netbeheer BV, Rotterdam, the Netherlands\\Email: \{s.zhan, j.morren, w.f.v.d.akker, a.e.v.d.molen, j.g.slootweg\}@tue.nl}}


\maketitle

\begin{abstract}
\input{0.abstract}
\end{abstract}


%
\IEEEpeerreviewmaketitle

\section{Introduction}
\input{1.intro}

\section{DER Coordination model}\label{sec:model}
\input{2.mathmodel}
\section{Case study}\label{sec:case}
\input{3.case}

\section{Conclusions}\label{sec:conclusion}
\input{4.conclusions}

\bibliography{reference}
\bibliographystyle{IEEEtran}

\end{document}

%% file: 0.abstract.tex
This study proposes a real-time distributed energy resource (DER) coordination model that can exploit flexibility from the DERs to solve voltage and overloading issues using both active and reactive power. The model considers time-coupling devices including electric vehicles and heat pumps by deviating as little as possible from their original schedules while prioritizing DERs with the most urgent demand using dynamic cost terms. The model does not require a multi-period setting or a multi-period-ahead forecast, which enables the model to alleviate the computational difficulty and enhance its applicability for distribution system operators to manage the grids in real time. A case study using a Dutch low-voltage grid assuming a 100\% penetration scenario of electric vehicles, heat pumps, and photovoltaics in the households validates that the proposed model can resolve the network issues while not affecting user comfort.

%% file: 1.intro.tex
Distribution grids are experiencing a transformation driven by decarbonization,  decentralization, and digitalization \cite{DISILVESTRE2018483}. On the supply side, distributed generators such as photovoltaics (PVs) are increasingly connected to the grids. While on the demand side, electric vehicles (EVs) and heat pumps (HPs) are deployed in residential households to replace their carbon-based counterparts. This new paradigm brings considerable operational issues to the distribution system operators (DSOs), such as cable overloading and voltage limit violation \cite{Nijhuis2015}.

Traditionally, the DSOs resolve these issues by reinforcing the grids. However, this process takes a long time and is costly. Fortunately, the recent roll-out of information and communication technology (ICT) infrastructures such as smart meters enables the DSOs to have two-way communication with the customers to make full use of flexibility from their distributed energy resources (DERs) to address these issues. 

Various studies in the literature have discussed using DER flexibility for grid management. While \cite{Jahangiri2013,Vergara2020,Mai2019} focused on autonomous droop-based control, many other studies developed communication-based control strategies. In \cite{Kotsalos2019}, a multi-period three-phase exact optimal power flow model was proposed to coordinate multiple DERs in the day-ahead timescale. The non-convex optimization problem was addressed with an interior-point method. In \cite{Gandhi2016}, a heuristic particle swarm optimization model was proposed to schedule reactive power from PVs and EVs to reduce network operation costs.

Concerning real-time operations, an optimal inverter dispatch problem was proposed in \cite{DallAnese2014}, which systematically determines active and reactive power setpoints of critical PV inverters. The binary PV-inverter selection variables were relaxed using a sparsity-promoting regularization approach, while the semidefinite relaxation was leveraged to characterize the non-linear power flow relations. In \cite{Kulmala2014}, a rule-based algorithm and an optimization-based one were both proposed to control the active and reactive power of DERs as well as transformer tap changer for voltage positioning. However, both studies do not include time-coupling devices such as EVs and HPs, which add substantial complexity to real-time control because control decisions at one time segment strongly affect subsequent time segments.

To consider time-coupling DERs for grid management, rolling horizon control (RHC) \cite{Benetti2015,Sabillon2018, Nazir2020} has gained increasing research interest in power system studies. In \cite{Benetti2015}, an updated 24-hour ahead forecast of power demand was leveraged for real-time control of EV charging. In \cite{Sabillon2018}, a one-day-ahead forecast of PV generation and EV demand was required for a dynamic scheduling model that controls on-load tap changers, PVs, EVs, and battery storage systems. While in \cite{Nazir2020}, a one-hour-ahead forecast of renewable generation and demand was used for the minute operation of inverter-interfaced energy storage in an unbalanced distribution feeder. The active power of energy storage obtained from the relaxed network model was fed into a nonlinear program that exactly characterizes the power flow relations for reactive power optimization. However, these RHC-based models bring additional computational difficulty by employing a multi-period optimization setting and require an accurate forecast for consecutive time segments, which might be unavailable on many occasions. At this point, a new research question emerges: \textit{how can various time-coupling DERs be coordinated in real-time network operations using only a one-step-ahead forecast?}

To address this research question, this paper proposes a convex optimization model which controls the active and reactive power of various DERs, including PVs, EVs, and HPs in real time for grid management, while avoiding a multi-period setting. The model aims to minimize the deviation from the customers' original schedules, while prioritizing DER users with the most urgent demand, which will likely ensure user satisfaction. The study \cite{Deilami2011} deals with a similar issue under the context of EV charging, which prioritizes EVs with the minimal impacts on the grids using, however, a rule-based procedure, which cannot guarantee optimality and did not include reactive power compensation from the EVs. 

In this respect, the main contribution of this study is to propose a convex optimization model which coordinates various time-coupling DERs using only a one-step-ahead forecast. The proposed model differs from other RHC-based models in the literature by avoiding a multi-period-ahead scheduling of time-coupling DERs. Instead, dynamic cost terms are employed to reflect DER users' urgentness in a single-period optimization setting to ensure user satisfaction. 


The rest of this paper is structured as follows. Section \ref{sec:model} introduces the convex DER coordination model and workflow in real-time grid operations. Section \ref{sec:case} presents a case study based on a Dutch low-voltage (LV) network, while Section \ref{sec:conclusion} draws conclusions and discusses future work.

%% file: 2.mathmodel.tex
\subsection{Operation Procedure}
\begin{figure}[tbp]
    \centering
    \includegraphics[width=.75\columnwidth]{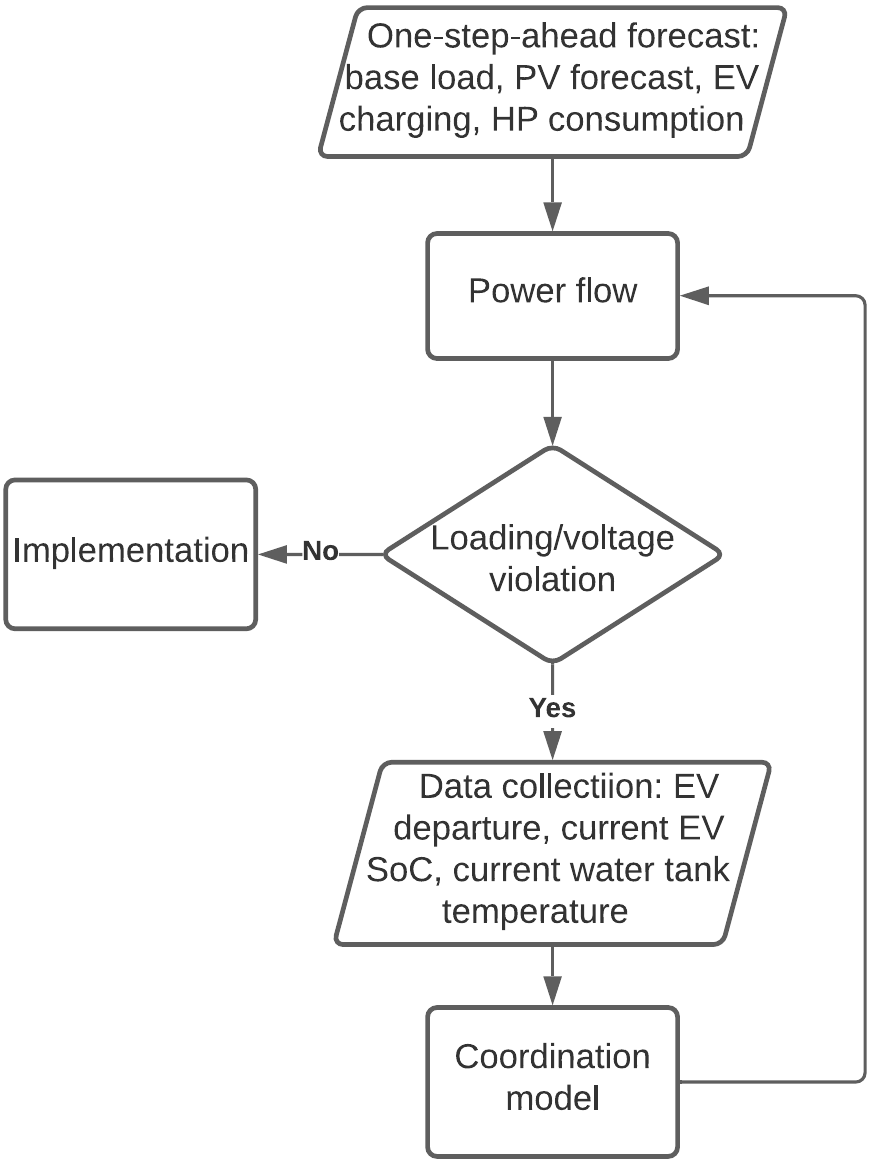}
    \caption{Real-time operation procedure}
    \label{fig:flowchart}
\end{figure}

This section introduces the real-time operation procedure for the DSO to manage the grid. PV capacity, EV power and energy capacity, HP capacity, water tank volume and temperature range are assumed known parameters to the DSO upon their installation. As shown in Fig. \ref{fig:flowchart}, before each operation step, the DSO collects power profiles from the customers via ICT infrastructures for the next time step, and runs a power flow program to evaluate the grid states. In case bus voltage limits or cable loading limits are violated, the DSO again collects data for model parameters and activates the coordination model to optimize the active and reactive power setpoints of various controllable DERs. It is assumed water tanks are used for domestic hot water heating and space heating. Their temperature is thus used as an important parameter to determine the urgentness of their power demand. Correspondingly, EVs' state of charge (SoC) and departure time are used to determine their urgentness. Finally, the optimized active and reactive power setpoints are re-fed into the power flow program for evaluation and grid state extraction considering that the proposed coordination model adopts relaxation for the power flow relations. Control actions are implemented afterwards.

\subsection{DER Coordination Model}
\begin{table}[tbp]
\begin{mdframed}
\input{nomenclature}
\printnomenclature
\end{mdframed}
\end{table}

This section presents the convex DER coordination model, which is implemented when grid limits are violated. Notations used are listed in the nomenclature. The model aims to restore grid limits by exploiting flexibility from various DERs, in which the decision variables include active and reactive power setpoints of EVs, HPs, and PVs connected to the grid. The objective function (\ref{obj}) minimizes the deviation from the DER users' original schedules collected from the customers, considering meanwhile losses in the grid. The cost terms are determined such that DER users with more urgent demand are assigned with higher values, which lowers the chance of their demand being curtailed. In this way, user satisfaction is more likely to be met. 

The constraints in the model are presented in the follows. The active power and reactive power injections into the grid from each household are defined with (\ref{pbalance}) and (\ref{qbalance}), respectively. The PVs' apparent power ratings are enforced in (\ref{pv1}), while (\ref{pv2}) reflects PV's power factor limits. Similar constraints for EVs are defined in (\ref{ev1}) and (\ref{ev2}). Fixed power factors for HPs are enforced in (\ref{hp}). Constraints (\ref{evcurtail}) and (\ref{pvcurtail}) respectively define active power curtailment for EVs and PVs. Constraints (\ref{hpcurtail1})-(\ref{hpcurtail3}) combined define the downward and upward regulation of active power for HPs.

The power flow relations are defined in (\ref{pflow})-(\ref{ilimits}), which adopts a second-order cone formulation \cite{Farivar2013}. Specifically, (\ref{pflow}) and (\ref{qflow}) define the active power and reactive power balance for each node, respectively. Voltage relation is enforced in (\ref{vrelations}), which is an exact formulation. The second-order cone relaxation is implemented in (\ref{socp}), which is relaxed from its equality form, defining apparent power flow for each cable. Constraints (\ref{vlimits}) and (\ref{ilimits}) define bus voltage limits and cable loading limits. Although there exist other convex relaxation models of the power flow relations such as semidefinite programming formulation \cite{Lavaei2012}, the second-order cone relaxation generally has a high accuracy and is exact under mild conditions for radial networks \cite{Farivar2013}, thus is widely used. 

\begin{subequations}
\begin{IEEEeqnarray}{rl} \label{obj}
&\underset{(p_j^{ev},p_j^{hp},p_j^{pv},q_j^{ev},q_j^{pv},\forall j \in \mathcal{H})}{\text{minimize}}\Big\{c^{loss}\sum_{\forall (i, j) \in \mathcal{E}} r_{i j} \ell_{i j} \nonumber\\&   +  \sum_{\forall j \in \mathcal{H}}(c_j^{ev} p_j^{ev\downarrow} + c_j^{pv} p_j^{pv\downarrow} + c_j^{hp\uparrow} p_j^{hp\uparrow}+ c_j^{hp\downarrow} p_j^{hp\downarrow} )\Big\} 
\end{IEEEeqnarray}
subject to
\begin{equation} \label{pbalance}
   p_j = p_j^{pv}-p_j^{load}-p_j^{ev}-p_j^{hp}  , \forall j \in \mathcal{H}
\end{equation}
\begin{equation} \label{qbalance}
   q_j = q_j^{pv}+q_j^{ev}-q_j^{hp}-q_j^{load} , \forall j\in \mathcal{H}
\end{equation}
\begin{equation} \label{pv1}
    (p_j^{pv})^2 +(q_j^{pv})^2 \leq (S_j^{pv})^2, \forall j\in \mathcal{H}
\end{equation}
\begin{equation} \label{pv2}
    - p_j^{pv} \tan \Phi^{pv,max} \leq q_j^{pv} \leq p_j^{pv} \tan \Phi^{pv,max}, \forall j\in \mathcal{H}
\end{equation}
\begin{equation} \label{ev1}
    (p_j^{ev})^2 +(q_j^{ev})^2 \leq (S_j^{ev})^2, \forall j\in \mathcal{H}
\end{equation}
\begin{equation} \label{ev2}
    - p_j^{ev} \tan \Phi^{ev,max} \leq q_j^{ev} \leq p_j^{ev} \tan \Phi^{ev,max}, \forall j\in \mathcal{H}
\end{equation}
\begin{equation} \label{hp}
    q_j^{hp} = p_j^{hp} \tan \Phi^{hp}, \forall j\in \mathcal{H}
\end{equation}
\begin{equation} \label{evcurtail}
    0 \leq p_j^{ev} \leq p_j^{ev,max}, p_j^{ev\downarrow} = p_j^{ev,max}-p_j^{ev}, \forall j\in \mathcal{H}
\end{equation}
\begin{equation}\label{pvcurtail}
    0 \leq p_j^{pv} \leq p_j^{pv,fore}, p_j^{pv\downarrow} = p_j^{pv,fore}-p_j^{pv}, \forall j\in \mathcal{H}
\end{equation}
\begin{equation}\label{hpcurtail1}
    0 \leq p_j^{hp} \leq p_j^{hp,max},  \forall j\in \mathcal{H}
\end{equation}
\begin{equation}\label{hpcurtail2}
    p_j^{hp\downarrow} \geq \max(0, p_j^{hp,set}-p_j^{hp}), \forall j\in \mathcal{H}
\end{equation}
\begin{equation}\label{hpcurtail3}
    p_j^{hp\uparrow} \geq \max(0, p_j^{hp}-p_j^{hp,set}), \forall j \in \mathcal{H}
\end{equation}
\begin{equation}\label{pflow}
p_{j}=\sum_{k: j \rightarrow k} P_{j k}-\sum_{i: i \rightarrow j}\left(P_{i j}-r_{i j} \ell_{i j}\right), \forall j \in \mathcal{N} 
\end{equation}
\begin{equation}\label{qflow}
q_{j}=\sum_{k: j \rightarrow k} Q_{j k}-\sum_{i: i \rightarrow j}\left(Q_{i j}-x_{i j} \ell_{i j}\right), \forall j\in \mathcal{N}
\end{equation}
\begin{IEEEeqnarray}{rl}\label{vrelations}
v_{j}=v_{i}-2\left(r_{i j} P_{i j}+x_{i j} Q_{i j}\right)+&\left(r_{i j}^{2}+x_{i j}^{2}\right) \ell_{i j},\\
&\forall(i, j) \in \mathcal{E} \nonumber
\end{IEEEeqnarray}
\begin{equation}\label{socp}
P_{i j}^{2}+Q_{i j}^{2} \leq \ell_{i j} v_i , \forall(i, j) \in \mathcal{E}
\end{equation}
\begin{equation}\label{vlimits}
   v_j^{min} \leq v_j \leq v_j^{max},  \forall j\in \mathcal{N}
\end{equation}
\begin{equation}\label{ilimits}
   0 \leq \ell_{i,j} \leq \ell_{i,j}^{max},
\forall(i, j) \in \mathcal{E}
\end{equation}
\end{subequations}

In summary, this model can restore grid limits by making full use of DER flexibility. It is noteworthy that this model requires only a one-step-ahead forecast, without the need for a multi-period setting and a multi-period-ahead forecast. The time-coupling DERs are coped with by deviating as little as possible from their original schedules while prioritizing DERs with the most urgent demand using dynamic cost terms.

%% file: nomenclature.tex
\nomenclature[W]{$i,j,k$}{Node index}
\nomenclature[W]{$\mathcal{H},\mathcal{N},\mathcal{E}$}{Household set, node set, cable set}

\nomenclature[P]{$c^{loss}$}{Loss cost, \texteuro/MWh}
\nomenclature[P]{$c_j^{ev},c_j^{pv}$}{EV/PV curtailment cost, \texteuro/MWh}
\nomenclature[P]{$c_j^{hp\uparrow},c_j^{hp\downarrow}$}{HP upward/downward regulation cost, \texteuro/MWh}
\nomenclature[P]{$r_{i,j},x_{i,j}$}{Resistance/reactance of cables, $\Omega$}
\nomenclature[P]{$p_j^{load},q_j^{load}$}{Active/reactive power of base load, MW/MVar}
\nomenclature[P]{$S_j^{pv},S_j^{ev}$}{Rated apparent power of PV/EV, MVA}
\nomenclature[P]{$\tan \Phi^{ev,max}$}{Maximum EV reactive/active power ratio}
\nomenclature[P]{$\tan \Phi^{pv,max}$}{Maximum PV reactive/active power ratio}
\nomenclature[P]{$\tan \Phi^{hp}$}{Fixed HP reactive/active power ratio}
\nomenclature[P]{$p_j^{ev,max}$}{Scheduled EV charging power, MW}
\nomenclature[P]{$p_j^{pv,fore}$}{PV generation forecast, MW}
\nomenclature[P]{$p_j^{hp,max}$}{HP capacity, MW}
\nomenclature[P]{$p_j^{hp,set}$}{Scheduled HP consumption, MW}
\nomenclature[P]{$v_j^{min},v_j^{max}$}{Node min/max voltage magnitude square, $(kV)^2$}
\nomenclature[P]{$\ell_j^{max}$}{Cable rating square, $(kA)^2$}

\nomenclature[V]{$v_j$}{Node voltage magnitude square, $(kV)^2$}
\nomenclature[V]{$\ell_j$}{Cable loading square, $(kA)^2$}
\nomenclature[V]{$p_j,q_j$}{Active/reactive power injection, MW/MVar}
\nomenclature[V]{$p_j^{ev},p_j^{hp},p_j^{pv}$}{EV/HP/PV active power, MW}
\nomenclature[V]{$q_j^{ev},q_j^{hp},q_j^{pv}$}{EV/HP/PV reactive power, MVar} 
\nomenclature[V]{$p_j^{ev\downarrow},p_j^{pv\downarrow}$}{EV/PV active power curtailment, MW}
\nomenclature[V]{$p_j^{hp\uparrow},p_j^{hp\downarrow}$}{HP upward/downward regulation, MW}
\nomenclature[V]{$P_{ij},Q_{ij}$}{Active/reactive power flow (i to j), MW/MVar}

%% file: 3.case.tex
\subsection{Case Description}
\begin{figure}[tbp]
    \centering
    \includegraphics[width=\linewidth]{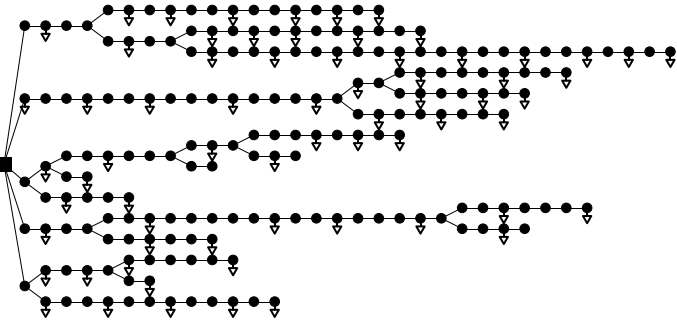}
    \caption{Dutch LV network topology}
    \label{fig:epse}
\end{figure}

This section presents a case study to evaluate the proposed procedure and DER coordination model for real-time grid operations. The simulation is performed on a Dutch LV network, shown in Fig. \ref{fig:epse}. A 100\% penetration scenario for EVs, PVs, and HPs is assumed for 67 connected residential households seeing the energy transition.

\begin{figure}[tbp]
    \centering
    \includegraphics[width=\linewidth]{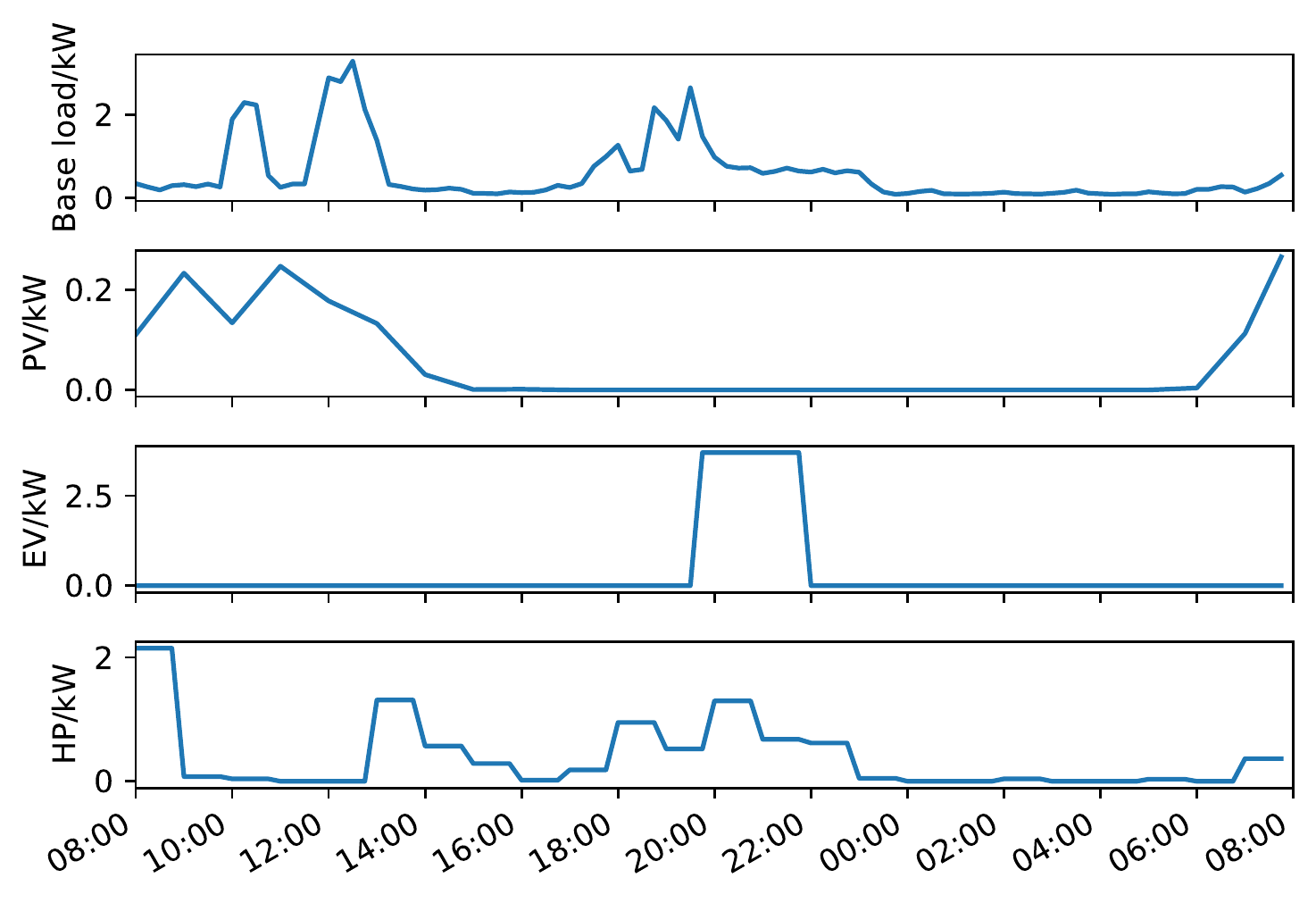}
    \caption{Base load, PV, EV, HP profiles for a household for the selected day, starting from 8:00 with a 15-minute resolution}
    \label{fig:datashow}
\end{figure}

At the first step, a power flow program is implemented for a year with a 15-minute resolution. From which, a winter day with the most significant under-voltage and overloading issues is selected to evaluate the proposed real-time operation procedure and coordination model. The base load profiles are based on anonymous smart meter data from a Dutch DSO, while the PV generation profile is calculated from an annual solar irradiance dataset from the Royal Netherlands Meteorological Institute assuming a 4.5kW PV installation for each household. The HP consumption profiles are based on anonymous smart meter gas usage data from a Dutch DSO, assuming 70\% of gas is used for domestic heating \cite{Ruhnau2019}. Literature \cite{Brunner2016, Sichilalu2017} can be referred to for HP and water tank models. For EV charging, the Monto Carlo simulation is performed by assuming normally distributed EV arrival time and a fixed electricity demand of 7.5kWh per day. A 230V, 16A charger with a 90\% efficiency is assumed for EV charging. For illustration purpose, base load, PV, EV, and HP profiles for a random household on the selected winter day is presented in Fig. \ref{fig:datashow}, where the coincidence of uncontrolled EV charging and HP consumption during nighttime is likely to result in under-voltage and overloading issues in the grid.

\subsection{Simulation Flowchart}
\begin{figure}[tbp]
    \centering
    \includegraphics[width=.75\linewidth]{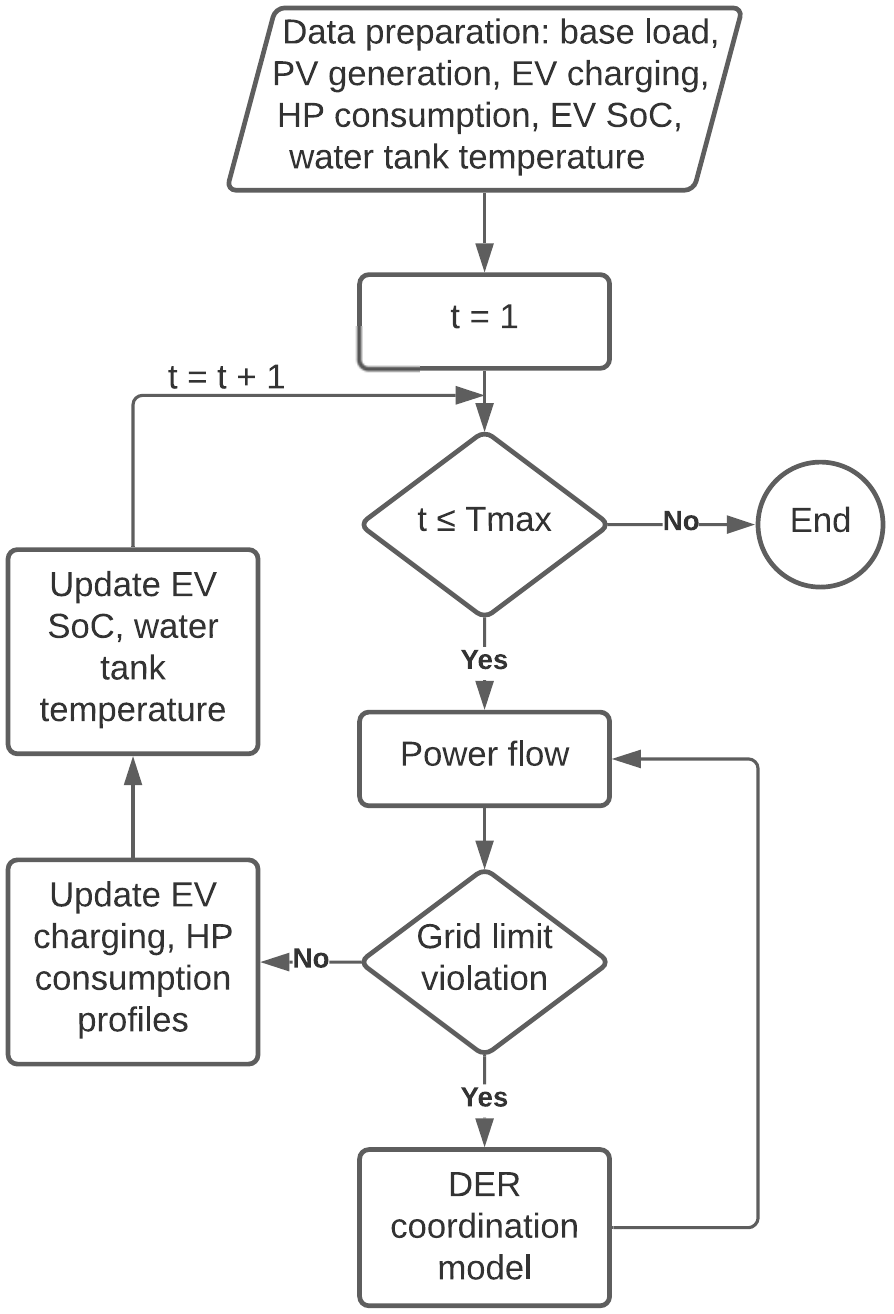}
    \caption{Simulation flowchart}
    \label{fig:simFlow}
\end{figure}

\begin{figure}[tbp]
    \centering
    \includegraphics[width=\linewidth]{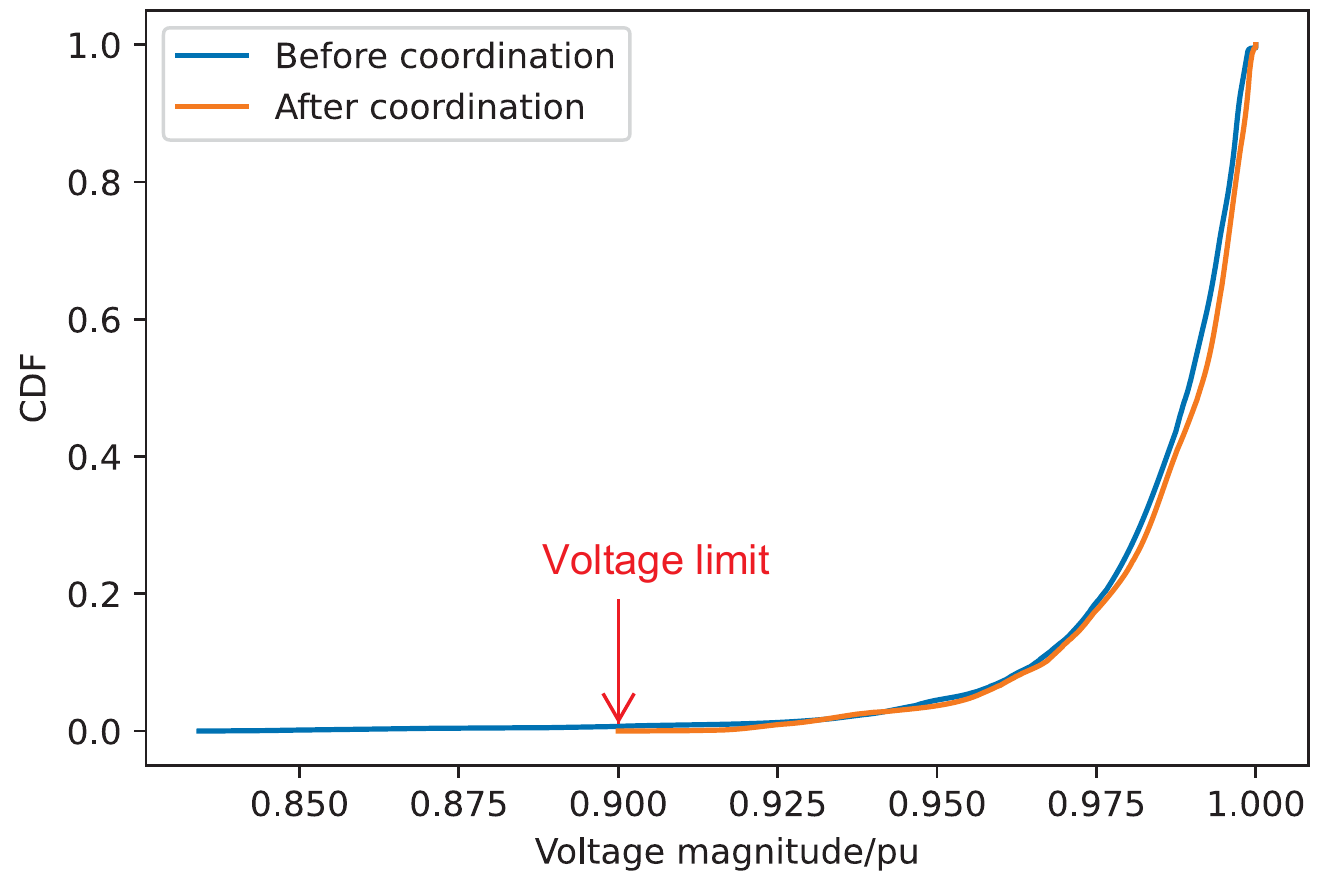}
    \caption{Bus voltage CDF}
    \label{fig:voltageCDF}
\end{figure}

\begin{figure}[tbp]
    \centering
    \includegraphics[width=\linewidth]{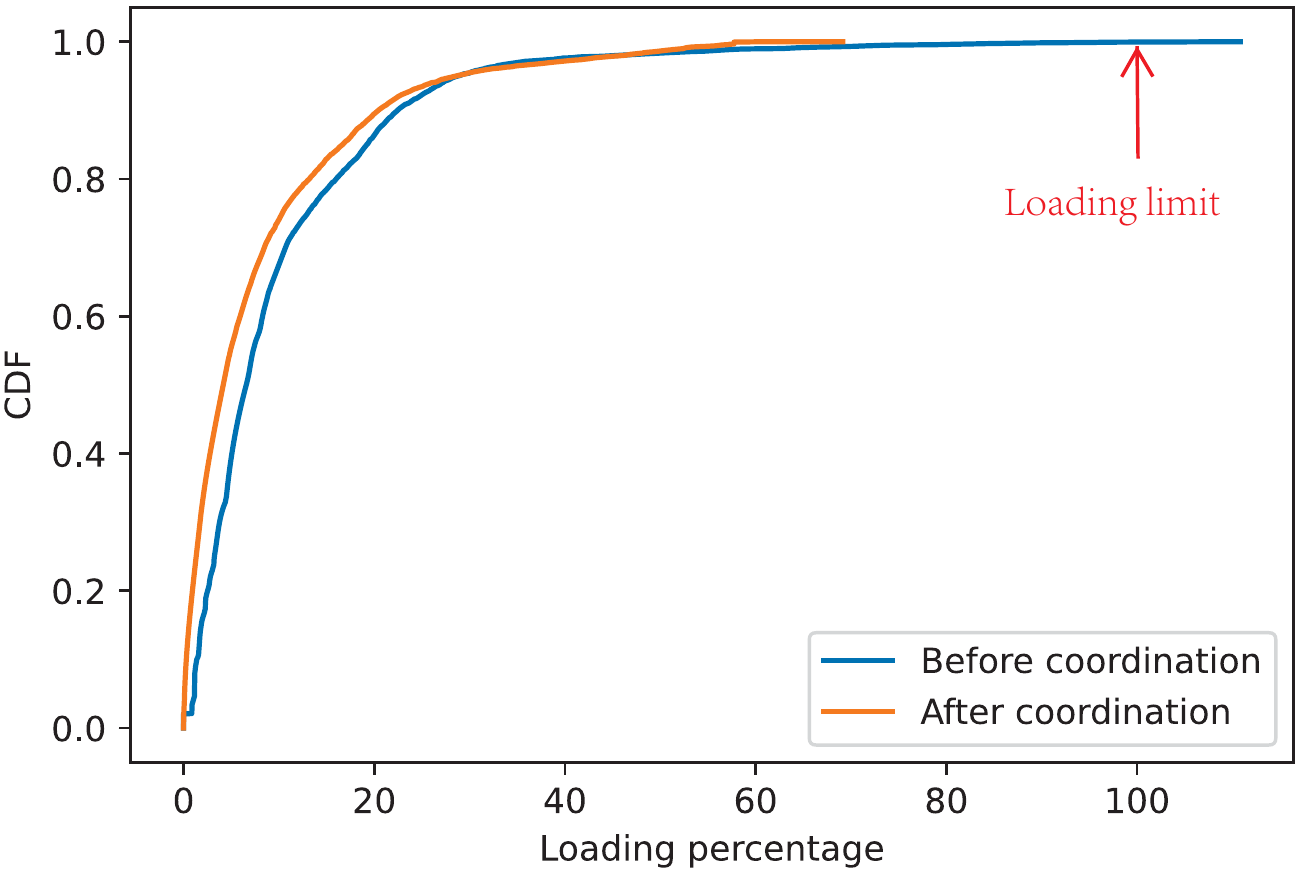}
    \caption{Cable loading CDF}
    \label{fig:loadingCDF}
\end{figure}

This section introduces the setting for the simulation on the selected day. The programs are implemented on a laptop with an Intel Core i5-9300H CPU running at 2.4GHz with 8GB RAM. The power flow program is implemented with Pandapower \cite{pandapower.2018}, while the optimization model is solved with Gurobi \cite{gurobi} with Python interface. The power flow program takes on average 1 second to solve, while the optimization model takes on average 4 seconds. It is noteworthy that a much higher resolution simulation could be performed under such short execution time, a 15-minute resolution simulation is still used considering the granularity of the available data.

The simulation process is described in Fig. \ref{fig:simFlow}, which starts with collecting various datasets as specified. At each simulation time step, a power flow program is implemented, followed by the DER coordination model in case any grid limit is violated. The optimized schedule from the coordination model is implemented, following which the EV charging and HP consumption profiles are updated for subsequent time slots to catch up on the deviation from the original schedules. For EVs, the charging process is extended in case the charging is curtailed. For HPs, the consumption is adjusted within their full capacity ranges to cancel out the deviation with the least possible delay. Finally, the EV state of charge (SoC) and water tank temperature are updated, which are used to calculate their corresponding cost terms in the DER coordination model and to evaluate user comfort.

\begin{table}[tbp]
    \centering
        \caption{Cost terms for the DER coordination model}
    \begin{tabular}{|c|c|c|} \hline
  \textbf{Cost term} & \textbf{Value} & \textbf{Note}\\ \hline
       $c^{loss}$  & 32\texteuro/MWh & Average day-ahead price \\ \hline
       $c^{pv}$  & 200\texteuro/MWh & Behind-the-meter\\\hline
       $c^{ev}$  & $\frac{c_0(1-SoC)}{(T_{max}-t+1)}$\texteuro/MWh & $c_0=1440, T_{max}=96$\\\hline
       $c^{hp\uparrow}$  & $\max(10,150\Delta T)$\texteuro/MWh & $\Delta T$ normalized \\\hline
       $c^{hp\downarrow}$  & $\max(10,-150\Delta T)$\texteuro/MWh & $\Delta T$ normalized \\\hline
    \end{tabular}
    \label{tab:cost}
\end{table}

For EVs, the cost terms are assumed to be functions of SoC and the remaining charging time. A unified departure time is assumed for the EVs, at the end of the simulation period. User satisfaction is determined by whether the EVs are fully charged before departure. For HPs, water tanks are assumed to supply hot water and space heating. A water tank of 1000L is assumed for households with no more than 5kW heating demand, while an additional 200L is assumed for each kW above 5kW \cite{Williams2012}. The water tanks are allowed for ${5}^{\circ} \mathrm{C}$ deviation up and down from the reference temperature, while not affecting user comfort \cite{Brunner2016, Sichilalu2017}. The cost terms are assumed as linear functions of water tank temperature deviation with a lower bound. User satisfaction is evaluated with respect to the water tank temperature limits. The used cost terms for this case study are listed in Table \ref{tab:cost}. It is noteworthy that in real-world applications, these cost terms can be derived through extensive simulation to guarantee the model's performance, or directly from end-users from a market perspective. With these model settings, the simulation results are presented in the following sections.

\subsection{Network Issues}

\begin{figure}[tbp]
    \centering
    \includegraphics[width=\linewidth]{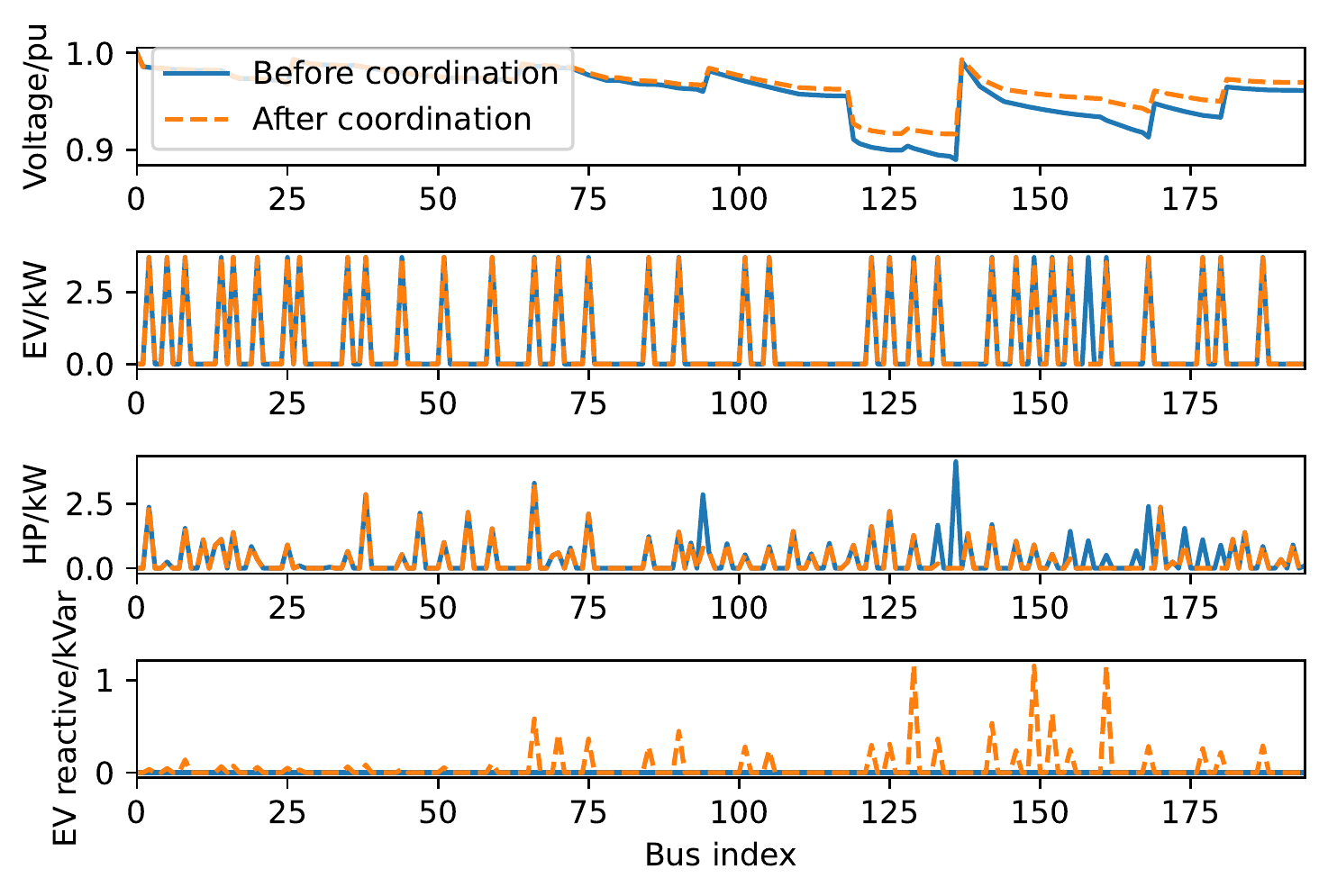}
    \caption{Voltage magnitude, EV/HP power and EV reactive power generation for a time slot}
    \label{fig:compare}
\end{figure}

As described above, the major issues envisioned in this case study are under-voltage and overloading, which have been confirmed in Fig. \ref{fig:voltageCDF} and Fig. \ref{fig:loadingCDF} respectively. The reported grid operation states are both from the power flow programs, instead of the optimization model. Using the DER coordination model, both issues have been completely resolved, where the bus voltage stays higher than 0.9 pu and the cable loading percentage is no more than 70\%. Fig. \ref{fig:compare} presents the voltage magnitude, EV and HP power demand, and EV reactive power generation for a time slot before and after implementation of the DER coordination model, which suggests that the restoration of voltage magnitude is attributed to curtailed EV and HP power demand, and EV reactive power compensation.

\subsection{User Satisfaction}

\begin{figure}[tbp]
    \centering
    \includegraphics[width=\linewidth]{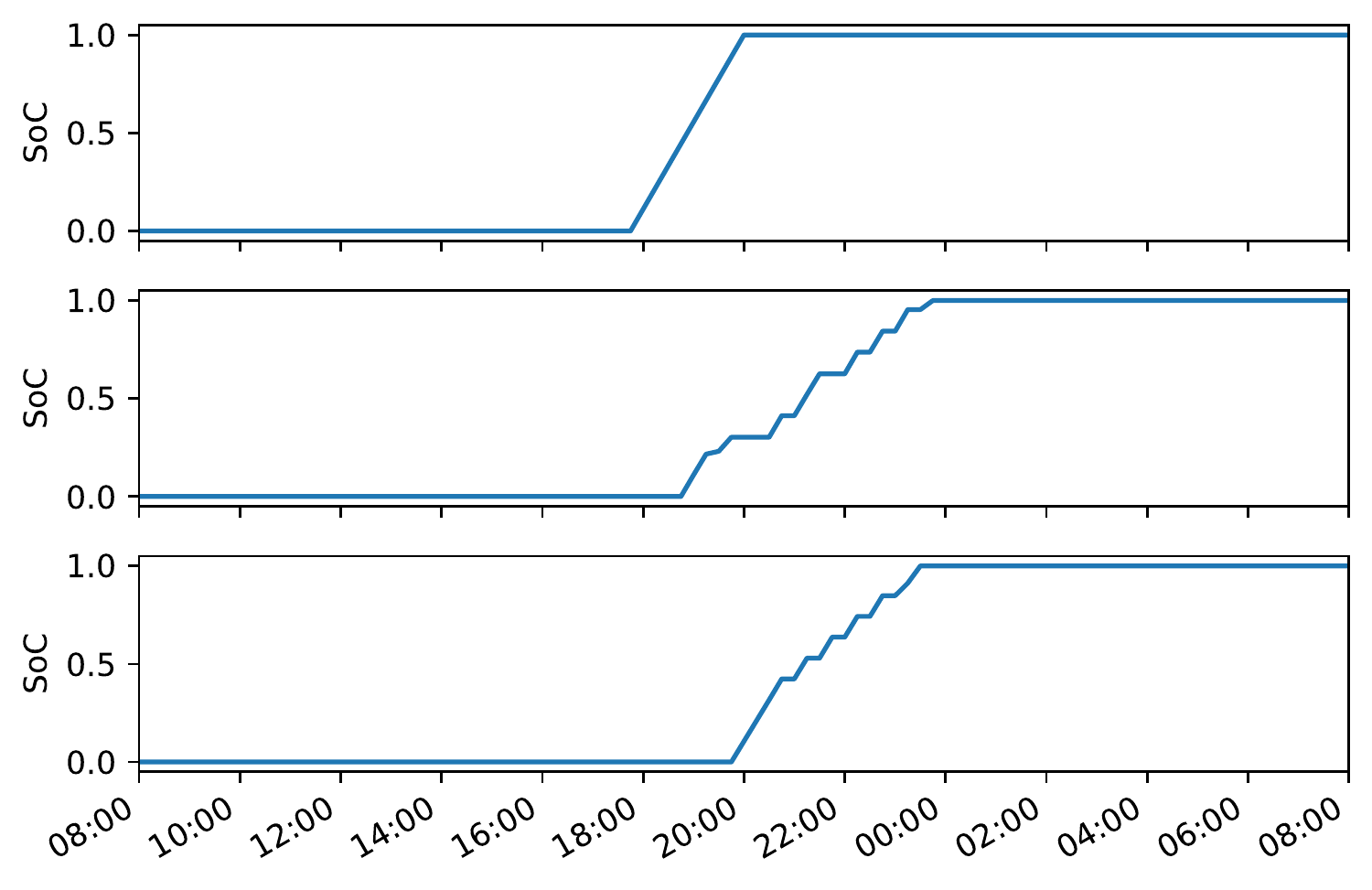}
    \caption{EV charging profiles for various households}
    \label{fig:evsatisfy}
\end{figure}

\begin{figure}[tbp]
    \centering
    \includegraphics[width=\linewidth]{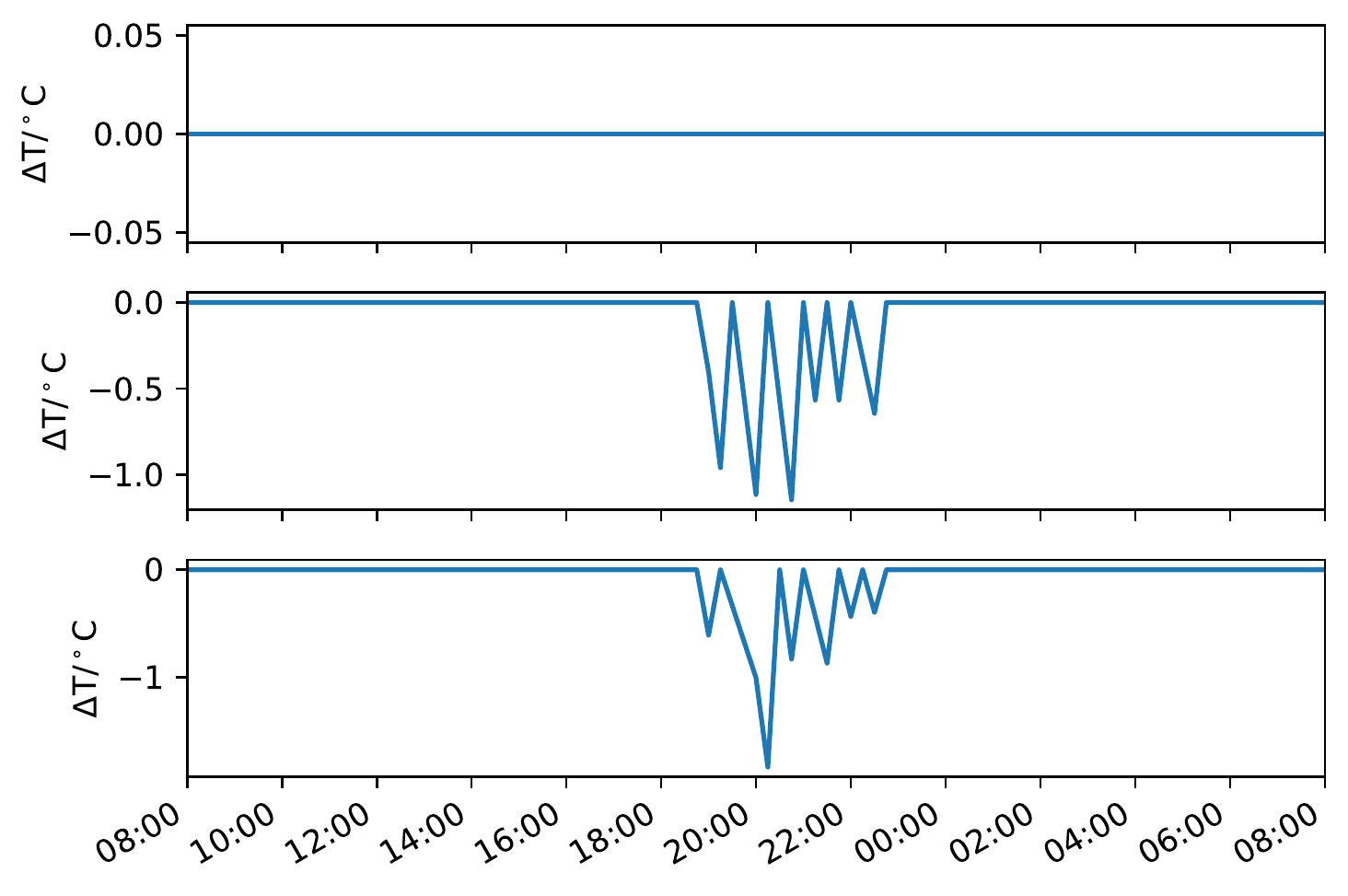}
    \caption{Water tank temperature deviation for various households}
    \label{fig:hpsatisfy}
\end{figure}

In this study, two indicators are used to evaluate user satisfaction: whether EVs have been fully charged, and whether the temperature deviation of water tanks has exceeded the upper and lower limits. Fig. \ref{fig:evsatisfy} shows 3 different charging profiles, where the first EV's charging process is not interrupted, while the rest two have been curtailed during the charging process. However, all EVs have been fully charged at the end of the simulation. Fig. \ref{fig:hpsatisfy} presents the water tank temperature changes for various households. The temperature has been quickly restored to the normal state after a HP curtailment occurs. It is shown that all water tanks have not violated the temperature limits (${5}^{\circ} \mathrm{C}$ up and down). Combining these results, it is demonstrated that user satisfaction has been kept using this DER coordination model.

\subsection{SOCP Relaxation Accuracy}
\begin{figure}[tbp]
    \centering
    \includegraphics[width=\linewidth]{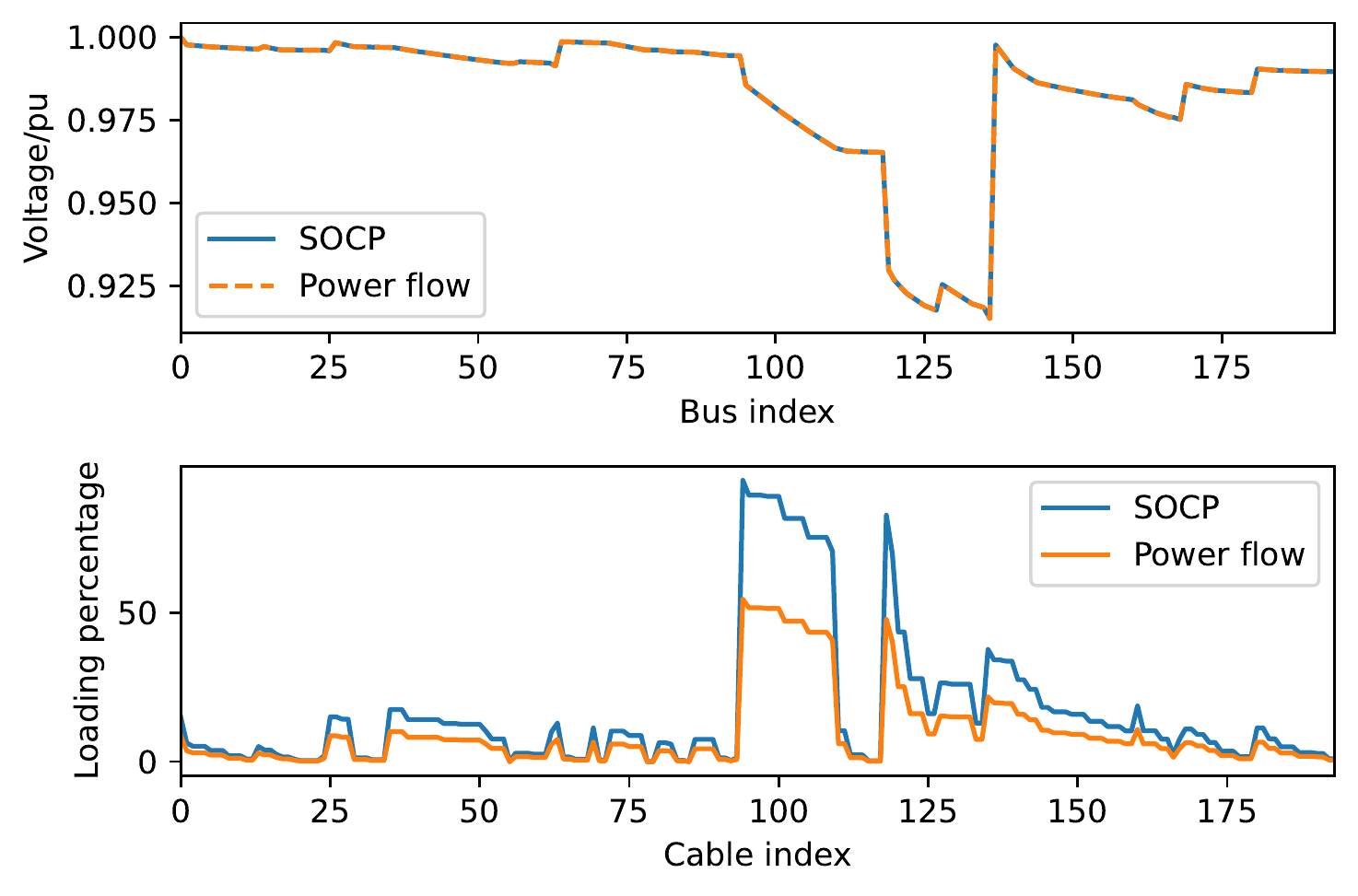}
    \caption{Voltage magnitude and cable loading from both SOCP relaxation model and power flow program}
    \label{fig:SOCP}
\end{figure}

This section deals with the accuracy of the adopted SOCP relaxation of the power flow relations. Fig. \ref{fig:SOCP} shows the voltage magnitude and cable loading for one time slot from the DER coordination model and the power flow program. It is seen that the voltage has been precisely captured by the SOCP model, while the cable loading has been over-estimated, due to the relaxation of the apparent power flow relation. However, an important finding is that the SOCP relaxation model ensures the feasibility of the optimized active and reactive power setpoints, which implies that when these control actions from the DER coordination model are implemented in the grid, it would not result in loading or voltage issues that are not captured in the SOCP model.

%% file: 4.conclusions.tex
A real-time DER coordination strategy is proposed in this paper to explore the flexibility from the DERs to resolve network issues using both active and reactive power. The model takes into account time-coupling devices including EVs and HPs without the need for a multi-period setting and a multi-period-ahead forecast, which significantly enhances its applicability for DSOs to manage the grids in real time. A case study using a Dutch LV grid assuming a 100\% penetration scenario of EVs, HPs, and PVs demonstrates the effectiveness of the model in solving network issues and ensuring user satisfaction. The ongoing energy transition around the world further outlines the importance of the model as more frequent network issues are expected to occur in the LV grids. Future work can start with a DER coordination model using three-phase networks, considering the phase unbalance. In addition, the aggregated effects on the medium-voltage and high-voltage grids from the proposed DER coordination model implemented in the LV grids can also be investigated.

%% file: main.bbl
\begin{thebibliography}{10}
\providecommand{\url}[1]{#1}
\csname url@samestyle\endcsname
\providecommand{\newblock}{\relax}
\providecommand{\bibinfo}[2]{#2}
\providecommand{\BIBentrySTDinterwordspacing}{\spaceskip=0pt\relax}
\providecommand{\BIBentryALTinterwordstretchfactor}{4}
\providecommand{\BIBentryALTinterwordspacing}{\spaceskip=\fontdimen2\font plus
\BIBentryALTinterwordstretchfactor\fontdimen3\font minus
  \fontdimen4\font\relax}
\providecommand{\BIBforeignlanguage}[2]{{%
\expandafter\ifx\csname l@#1\endcsname\relax
\typeout{** WARNING: IEEEtran.bst: No hyphenation pattern has been}%
\typeout{** loaded for the language `#1'. Using the pattern for}%
\typeout{** the default language instead.}%
\else
\language=\csname l@#1\endcsname
\fi
#2}}
\providecommand{\BIBdecl}{\relax}
\BIBdecl

\bibitem{DISILVESTRE2018483}
M.~L. {Di Silvestre}, S.~Favuzza, E.~{Riva Sanseverino}, and G.~Zizzo, ``How
  decarbonization, digitalization and decentralization are changing key power
  infrastructures,'' \emph{Renewable and Sustainable Energy Reviews}, vol.~93,
  pp. 483--498, 2018.

\bibitem{Nijhuis2015}
M.~Nijhuis, M.~Gibescu, and J.~F. Cobben, ``{Assessment of the impacts of the
  renewable energy and ICT driven energy transition on distribution
  networks},'' \emph{Renewable and Sustainable Energy Reviews}, vol.~52, pp.
  1003--1014, aug 2015.

\bibitem{Jahangiri2013}
P.~Jahangiri and D.~C. Aliprantis, ``{Distributed Volt/VAr control by PV
  inverters},'' \emph{IEEE Transactions on Power Systems}, vol.~28, no.~3, pp.
  3429--3439, 2013.

\bibitem{Vergara2020}
P.~P. Vergara, M.~Salazar, T.~T. Mai, P.~H. Nguyen, and H.~Slootweg, ``{A
  comprehensive assessment of PV inverters operating with droop control for
  overvoltage mitigation in LV distribution networks},'' \emph{Renewable
  Energy}, vol. 159, pp. 172--183, 2020.

\bibitem{Mai2019}
T.~T. Mai, N.~A. Haque, H.~T. Vo, and P.~H. Nguyen, ``{Coordinated active and
  reactive power control for overvoltage mitigation in physical LV
  microgrids},'' \emph{The Journal of Engineering}, vol. 2019, no.~18, pp.
  5007--5011, 2019.

\bibitem{Kotsalos2019}
K.~Kotsalos, I.~Miranda, N.~Silva, and H.~Leite, ``{A horizon optimization
  control framework for the coordinated operation of multiple distributed
  energy resources in low voltage distribution networks},'' \emph{Energies},
  vol.~12, no.~6, pp. 1--27, 2019.

\bibitem{Gandhi2016}
O.~Gandhi, W.~Zhang, C.~D. Rodriguez-Gallegos, D.~Srinivasan, and T.~Reindl,
  ``{Continuous optimization of reactive power from PV and EV in distribution
  system},'' \emph{IEEE PES Innovative Smart Grid Technologies Conference
  Europe}, pp. 281--287, 2016.

\bibitem{DallAnese2014}
E.~Dall'Anese, S.~V. Dhople, and G.~B. Giannakis, ``{Optimal dispatch of
  photovoltaic inverters in residential distribution systems},'' \emph{IEEE
  Transactions on Sustainable Energy}, vol.~29, no.~4, pp. 957--967, 2014.

\bibitem{Kulmala2014}
A.~Kulmala and S.~Repo, ``{Coordinated Voltage Control in Distribution Networks
  Including Several Distributed Energy Resources},'' \emph{IEEE Transactions on
  Smart Grid}, vol.~5, no.~4, pp. 2010--2020, 2014.

\bibitem{Benetti2015}
G.~Benetti, M.~Delfanti, T.~Facchinetti, D.~Falabretti, and M.~Merlo,
  ``{Real-Time Modeling and Control of Electric Vehicles Charging Processes},''
  \emph{IEEE Transactions on Smart Grid}, vol.~6, no.~3, pp. 1375--1385, 2015.

\bibitem{Sabillon2018}
C.~Sabillon, J.~F. Franco, M.~J. Rider, and R.~Romero, ``{Joint optimal
  operation of photovoltaic units and electric vehicles in residential networks
  with storage systems: A dynamic scheduling method},'' \emph{International
  Journal of Electrical Power and Energy Systems}, vol. 103, no. March, pp.
  136--145, 2018.

\bibitem{Nazir2020}
N.~Nazir, P.~Racherla, and M.~Almassalkhi, ``{Optimal Multi-Period Dispatch of
  Distributed Energy Resources in Unbalanced Distribution Feeders},''
  \emph{IEEE Transactions on Power Systems}, vol.~35, no.~4, pp. 2683--2692,
  2020.

\bibitem{Deilami2011}
S.~Deilami, A.~S. Masoum, P.~S. Moses, and M.~A. Masoum, ``{Real-time
  coordination of plug-in electric vehicle charging in smart grids to minimize
  power losses and improve voltage profile},'' \emph{IEEE Transactions on Smart
  Grid}, vol.~2, no.~3, pp. 456--467, 2011.

\bibitem{Farivar2013}
M.~Farivar and S.~H. Low, ``{Branch flow model: Relaxations and
  convexification-part i},'' \emph{IEEE Transactions on Power Systems},
  vol.~28, no.~3, pp. 2554--2564, 2013.

\bibitem{Lavaei2012}
J.~Lavaei and S.~H. Low, ``{Zero duality gap in optimal power flow problem},''
  \emph{IEEE Transactions on Power Systems}, vol.~27, no.~1, pp. 92--107, 2012.

\bibitem{Ruhnau2019}
O.~Ruhnau, L.~Hirth, and A.~Praktiknjo, ``{Time series of heat demand and heat
  pump efficiency for energy system modeling},'' \emph{Scientific Data},
  vol.~6, no.~1, pp. 1--10, 2019.

\bibitem{Brunner2016}
M.~Brunner, K.~Rudion, and S.~Tenbohlen, ``{PV curtailment reduction with smart
  homes and heat pumps},'' \emph{IEEE International Energy Conference,
  ENERGYCON}, pp. 1--6, 2016.

\bibitem{Sichilalu2017}
S.~Sichilalu, T.~Mathaba, and X.~Xia, ``{Optimal control of a wind–PV-hybrid
  powered heat pump water heater},'' \emph{Applied Energy}, vol. 185, pp.
  1173--1184, 2017.

\bibitem{pandapower.2018}
L.~Thurner, A.~Scheidler, F.~Schafer, J.~H. Menke, J.~Dollichon, F.~Meier,
  S.~Meinecke, and M.~Braun, ``pandapower - an open source python tool for
  convenient modeling, analysis and optimization of electric power systems,''
  \emph{IEEE Transactions on Power Systems}, 2018.

\bibitem{gurobi}
\BIBentryALTinterwordspacing
{Gurobi Optimization, LLC}, ``Gurobi optimizer reference manual,'' 2021.
  [Online]. Available: \url{http://www.gurobi.com}
\BIBentrySTDinterwordspacing

\bibitem{Williams2012}
C.~J. Williams, J.~O. Binder, and T.~Kelm, ``{Demand side management through
  heat pumps, thermal storage and battery storage to increase local
  self-consumption and grid compatibility of PV systems},'' \emph{IEEE PES
  Innovative Smart Grid Technologies Conference Europe}, pp. 1--6, 2012.

\end{thebibliography}
